\documentclass{article}
\usepackage{amsmath,amssymb,amsthm,comment,mathtools}
\usepackage[usenames]{color}
\usepackage{hyperref}
\usepackage{url}

\newtheorem{theorem}{Theorem}
\newtheorem{corollary}{Corollary}
\newtheorem{proposition}{Proposition}
\newtheorem{protocol}{Protocol}

\theoremstyle{definition}

\newtheorem{remark}[theorem]{Remark}

\newcommand{\C}{\mathbb C}

\renewcommand\H{\ensuremath{\mathcal H}}
\newcommand\K{\ensuremath{\mathcal K}}

\renewcommand\phi{\varphi}
\newcommand\Prob{\ensuremath{\text{Prob}}}

\newcommand{\R}{\ensuremath{{\mathbb R}}}
\newcommand\W{\ensuremath{\mathbf W}}

\title{Negative probabilities}
\author{Yuri Gurevich%
  \thanks{Partially supported by the US Army Research Office under W911NF-20-1-0297.}\\
  {\small Computer Science \&\ Engineering}\\[-1mm]
  {\small University of Michigan}\\[-1mm]
  {\small Ann Arbor, Michigan, U.S.A.}
  \and
  Vladimir Vovk%
  \thanks{Partially supported by Amazon, Astra Zeneca, and Stena Line.}\\
  {\small Dept.\ of Computer Science}\\[-1mm]
  {\small Royal Holloway, University of London}\\[-1mm]
  {\small Egham, Surrey, UK}}
\date{}

\begin{document}
\maketitle

\begin{abstract}
We explain, on the example of Wigner's quasiprobability distribution, how negative probabilities may be used in the foundations of probability.
\end{abstract}

\section{Introduction}

Negative probabilities may appear devoid of any empirical meaning.
The frequentist semantics does not apply to them.
``It is absurd to talk about an urn containing $-17$ red balls" \cite[page~148]{Pitowsky}.
Yet negative probabilities are usefully employed in quantum physics. The situation reminds us of the introduction of imaginary numbers in the theory of algebraic equations, even though the standard semantics, according to which numbers are quantities, does not apply to imaginary numbers.

In this note we illustrate that negative probabilities may be usefully employed in the game-theoretic approach to probabilities \cite{SV01,SV19}.

\section{Quantum measurement: statistical testing view}

We work in the framework of the most standard formalization of (non-relativistic) quantum mechanics. It was originally proposed by John von Neumann \cite{Neumann1929,Neumann1932}. A modern, mathematically rigorous exposition is found in the book \cite{Hall}.

Like any physical system, a quantum system $Q$ has a state space. The state space of $Q$ is a Hilbert space \H. The states of $Q$ are represented by unit vectors in \H.

For simplicity of exposition, we consider a quantum system $Q$ of one particle $\Pi$ moving in one dimension, though all our results generalize to more (but finitely many) particles moving in more (but finitely many) dimensions.
The state space of our quantum system $Q$ is the Hilbert space $L^2(\R)$ of square integrable functions $f:\R\to\C$
with the inner product of $f,g\in L^2(\R)$ given by the Lebesgue integral $\displaystyle\int f^*(t) g(t) dt$. In this note, by default, the integrals are from $-\infty$ to $+\infty$.

Consider physical properties of $Q$, like the position of particle $\Pi$, which take real values and can be measured. Such a physical property is represented by a self-adjoint operator $A$ over $L^2(\R)$. It is common to speak about measuring $A$ itself and to call $A$ an observable.
The result of the measurement of (the physical property represented by) $A$ in a given state $\psi$ of $Q$ is determined probabilistically. The probability that the result lies in real interval $(p,q]$ is given by formula
\begin{equation}\label{star}
 \Prob_A(p,q] = \left\|(E_q - E_p)\psi\right\|^2 \tag{$*$}
\end{equation}
where $\{E_r: r\in\R\}$ is the spectral resolution of the identity for $A$, whose existence follows from the spectral theorem for linear operators in Hilbert spaces proved originally by von Neumann \cite{Neumann1929}; a modern treatment of the issue is found in \cite[\S10]{Hall}.

We represent quantum measurement as the following protocol of interaction between four players, Experimenter, Quantum Mechanics, Skeptic, and Reality. (The identities of Experimenter and Reality are not essential to us; we could combine them in one super player, World.)

\begin{protocol}[Quantum measurement]\mbox{}\label{prot:1}
\rm
\begin{itemize}
\item[] $\K_0 := 1$.
\item[] FOR $n=1,2,\dots$,
  \begin{enumerate}
  \item Experimenter (prepares and) announces a state $\psi_n$ of system $Q$\\ and an observable $A_n$ to be measured in state $\psi_n$.
  \item Quantum Mechanics announces a probability distribution $\mu_n$ on \R.
  \item Skeptic announces a measurable function $f_n\in[0,\infty]^\R$\\
        such that $\displaystyle\int\! f_n d \mu_n = 1$.
  \item Reality announces the result $r_n\in\R$ of the measurement of observable $A_n$ in state $\psi_n$.
  \item $\K_n := \K_{n-1} f_n (r_n)$.
  \end{enumerate}
\end{itemize}
\end{protocol}

\bigskip
Here Experimenter and Reality are free agents, who do not have to follow any strategy, deterministic or probabilistic.
The strategy of Quantum Mechanics is given by formula \eqref{star} where, at stage $n$, the observable is $A_n$ so that $\mu_n = \Prob_{A_n}$.
The goal of Skeptic is to test Quantum Mechanics, and we will be interested in testing strategies for Skeptic.

Protocol~\ref{prot:1} is a simplified version of the protocols
given in \cite[\S10.6]{SV19} and \cite[\S8.4]{SV01},
which also involve the deterministic development of state $\psi$ governed by the Schr\"odinger equation.
Here is one corollary of game-theoretic limit theorems
(cf.\ \cite[Corollary 10.14]{SV19}).

\begin{corollary}\label{cor:1}
Let $F:\R\to\R$ be a bounded measurable function.
Skeptic can force the event
  \begin{equation*}
    \lim_{N\to\infty}
    \frac1N
    \sum_{n=1}^N
    \left(
      F(r_n) - \int\! F d\mu_n
    \right)
    = 0,
  \end{equation*}
in the sense of having a strategy ensuring $\K_n\to\infty$ whenever the equality fails.
\end{corollary}

This corollary (a law of large numbers) is unusual in that it lies outside Kolmogorov's framework for probability. The reason for that phenomenon is that Experimenter does not have to follow any strategy.
Of course, real-world experimenters usually follow deterministic or probabilistic testing strategies, which brings us into Kolmogorov's framework.

\section{Wigner's quasiprobability distribution}

Recall that our quantum system $Q$ is a particle $\Pi$ moving in one dimension.
For technical reasons, we assume that Experimenter only prepares $Q$ in states $\psi$ which are smooth and compactly supported on $\R$; such states $\psi$ will be called \emph{nice}. Nice states are everywhere dense in $L^2(\R)$. Below, by default, states of $Q$ are nice.

Two important physical properties of $Q$ are the position $x$ and momentum $p$ of particle $\Pi$.
According to quantum mechanics, they are represented by self-adjoint operators
\[
  (X\psi)(x) := x\psi(x)
  \quad \text{and} \quad
  (P\psi)(x) := -i\hbar \frac{d\psi}{dx}(x)
\]
respectively where $\hbar$ is a real constant, the so-called reduced Planck constant.

The uncertainty principle of quantum mechanics asserts a limit to the precision with which position $x$ and momentum $p$ can be determined simultaneously in a given state $\psi$, even if $\psi$ is nice.
You can know the distribution of $x$ and that of $p$,
but there is no joint probability distribution of $x,p$ with the correct marginal distributions of $x$ and $p$.

The situation changes if one allows negative probabilities. To address the issue, we need the following definitions. A \emph{quasiprobability distribution} $\mu$ on a measurable space $(\Omega,\Sigma)$ is a real-valued, countably additive function on the measurable sets such that $\mu(\Omega)=1$. A \emph{quasiprobability density function} for a given quasiprobability distribution is the obvious generalization of a probability density function for a given probability distribution.

\begin{remark}
Qasiprobability distributions are special signed probability measures and are also known as \emph{signed probability distributions}. One may worry whether countable additivity makes sense in signed probability spaces, but it does \cite[\S2.1]{G145}.
\end{remark}

In a 1932 paper \cite{Wigner}, Eugene Wigner exhibited a function
\begin{equation*}
  W_{\psi}(x,p) := \frac{1}{2\pi}
  \int \psi^*\!\! \left(x + \frac{\beta\hbar}{2}\right)
  \psi\! \left(x - \frac{\beta\hbar}{2}\right)
  e^{i\beta p} d\beta
\end{equation*}
where $\psi$ is an arbitrary unit vector in $L^2(\R)$. It is easy to check that all values of Wigner's function $W_{\psi}(x,p)$ are real, but some values may be negative.
In any nice state $\psi$, $W_\psi$ gives rise to a unique quasiprobability distribution $\W_{\psi}$, \emph{Wigner's quasiprobability distribution}, for which it is a quasiprobability density function.

\begin{remark}
The Wigner function $W_{\psi}$ is also known as the Wigner-Ville function because it was introduced in 1948 by Jean-Andr\'e Ville in the context of signal processing \cite{Flandrin,Ville}. Signal processing is beyond the scope of this paper.
\end{remark}

For any real numbers $a,b$, the physical property $z = ax + bp$ of $Q$ is represented by the self-adjoint operator $Z = aX + bP$.
In any nice state $\psi$ of quantum system $Q$, let $w^z_\psi$, or $w^{a,b}_\psi$, be the probability distribution $\Prob_Z$ given by formula \eqref{star} with $A = Z$. (The alternative notation $w^{a,b}_\psi$ is more explicit, but we will use the more succinct notation $w^z_\psi$.)

The following proposition was presented in \cite{Bertrand} and rigorously proved in \cite{G145}.

\begin{proposition}\label{prop}
In every nice state $\psi$, Wigner's quasiprobability distribution $\W_{\psi}$ is the unique quasiprobability distribution on $\R^2$ whose image, under any linear mapping
$\displaystyle (x,p) \mapsto z = ax + bp $,
is exactly $w^z_\psi$.
\end{proposition}

Although $\W_{\psi}$ often has negative values, its images $w^z_\psi$ are genuine nonnegative probability distributions.
The proposition allows us to refine Protocol~\ref{prot:1} as follows.

\begin{protocol}[Wigner-style quantum measurement]\mbox{}
\label{prot:2}\rm
\begin{itemize}
\item[] $\K_0 := 1$.
\item[] FOR $n=1,2,\dots$,
  \begin{enumerate}
  \item Experimenter prepares and announces a nice state $\psi_n$.
  \item Quantum Mechanics announces a quasiprobability distribution, namely $\W_{\psi_n}$.
  \item Experimenter chooses a physical property $z_n = a_n x + b_n p$ and announces the observable $Z_n = a_n X + b_n P$.
  \item Skeptic announces a measurable function $f_n\in[0,\infty]^\R$ such that
    $\displaystyle \int f_n\, d w^{z_n}_{\psi_n} = 1$.
  \item Reality announces the result $r_n\in\R$ of the measurement of observable $Z_n$ in state $\psi_n$.
  \item $\K_n := \K_{n-1} f_n (r_n)$.
  \end{enumerate}
\end{itemize}
\end{protocol}

\bigskip
Protocol~\ref{prot:2} shows how we can test Wigner's quasiprobability distribution.
Notice that Skeptic gambles only against nonnegative probability distributions $w^{z_n}_{\psi_n}$.
This is a coherent testing protocol in the sense of \cite{SV01,SV19}.
Recall that each probability distribution  $w^{z_n}_{\psi_n}$ is an image of quasiprobability distribution $\W_{\psi_n}$;
negative probabilities are used only to generate nonnegative probabilities which are tested as usual.

\begin{corollary}\label{cor:2}
  Let $F:\R\to\R$ be a bounded measurable function.
  Skeptic can force the event
  \begin{equation*}
    \lim_{N\to\infty}
    \frac1N
    \sum_{n=1}^N
    \left(
      F(r_n)
      -
      \int\!\!
      \int
      F(a_n x + b_n p)
      W_{\psi_n}(x,p)
      d x \, d p
    \right)
    = 0
  \end{equation*}
in the sense of having a strategy ensuring $\K_n\to\infty$ whenever the equality fails.
\end{corollary}

Corollary~\ref{cor:2} is stronger than Corollary~\ref{cor:1} in the sense that, in Protocol~\ref{prot:2}, Quantum Mechanics makes its choice \emph{before} Experimenter announces an observable. By Proposition~\ref{prop}, this is impossible to achieve without negative probabilities.

\section{Conclusion}

Wigner's function is a simple and concise description of probabilistic predictions
for a wide range of observables.
It can be tested using the usual approach of game-theoretic probability.
The function has found useful applications in physics \cite{G145} and signal processing \cite{Flandrin},
and we expect that the role of quasiprobability distributions will only grow both in practice
and in the foundations of probability and statistics.

\subsection*{Acknowledgement}
We thank Andreas Blass for useful comments.

\end{document}